\documentclass[conference]{IEEEtran}
\IEEEoverridecommandlockouts
\usepackage{cite}
\usepackage{amsmath,amssymb,amsfonts}
\usepackage{algorithmic}
\usepackage{graphicx}
\usepackage{subcaption}
\usepackage{textcomp}
\usepackage{xcolor}

\usepackage{amsthm}
\usepackage{setspace}
\usepackage{lipsum}
\theoremstyle{plain}

\theoremstyle{definition}

\theoremstyle{remark}

\usepackage{footnote}
\theoremstyle{plain}

\usepackage{fancybox,framed}

\def\BibTeX{{\rm B\kern-.05em{\sc i\kern-.025em b}\kern-.08em
    T\kern-.1667em\lower.7ex\hbox{E}\kern-.125emX}}
\begin{document}

\newcommand{\nt}[1]{\textcolor{red}{\textbf{[#1]}}}

\newcommand{\tcmajor}[1]{{{#1}}}
\newcommand{\chl}[1]{{\color{red}{#1}}}

\title{Distribution-Aware GMD Transceiver Design for Probabilistic Shaping in MIMO}
% \\
% {\footnotesize \textsuperscript{*}Note: Sub-titles are not captured in Xplore and
% should not be used}
% \thanks{Identify applicable funding agency here. If none, delete this.}
% }

\author{Tzu-Hsuan~{Chou}, Chih-Hao~{Liu}, and Jing~{Jiang}\\
% \IEEEauthorblockA{\textit{Qualcomm Technologies, Inc., San Diego, CA, USA} \\
% % \textit{name of organization (of Aff.)}\\
% % City, Country \\
% \{tzuhchou, chihliul\}@qti.qualcomm.com}
		\thanks{Copyright (c) 2026 IEEE. Personal use of this material is permitted. However, permission to use this material for any other purposes must be obtained from the IEEE by sending a request to pubs-permission@ieee.org.}
        \thanks{This work was supported by Qualcomm Wireless Research. }
		\thanks{T.-H. Chou, C.-H. Liu and J. Jiang are with Qualcomm, Inc., San Diego, CA 92121 USA; (e-mail: \{tzuhchou, chihliul, jingj\}@qti.qualcomm.com).}
        % \thanks{C.-H. Huang is with the Department of Electrical Engineering, National Taiwan University, Taipei, Taiwan; emails: }
		%	\end{singlespace} 
		\vspace{-7mm}
	}

% \author{\IEEEauthorblockN{1\textsuperscript{st} Given Name Surname}
% \IEEEauthorblockA{\textit{dept. name of organization (of Aff.)} \\
% \textit{name of organization (of Aff.)}\\
% City, Country \\
% email address or ORCID}
% \and
% \IEEEauthorblockN{2\textsuperscript{nd} Given Name Surname}
% \IEEEauthorblockA{\textit{dept. name of organization (of Aff.)} \\
% \textit{name of organization (of Aff.)}\\
% City, Country \\
% email address or ORCID}
% \and
% \IEEEauthorblockN{3\textsuperscript{rd} Given Name Surname}
% \IEEEauthorblockA{\textit{dept. name of organization (of Aff.)} \\
% \textit{name of organization (of Aff.)}\\
% City, Country \\
% email address or ORCID}
% \and
% \IEEEauthorblockN{4\textsuperscript{th} Given Name Surname}
% \IEEEauthorblockA{\textit{dept. name of organization (of Aff.)} \\
% \textit{name of organization (of Aff.)}\\
% City, Country \\
% email address or ORCID}
% \and
% \IEEEauthorblockN{5\textsuperscript{th} Given Name Surname}
% \IEEEauthorblockA{\textit{dept. name of organization (of Aff.)} \\
% \textit{name of organization (of Aff.)}\\
% City, Country \\
% email address or ORCID}
% \and
% \IEEEauthorblockN{6\textsuperscript{th} Given Name Surname}
% \IEEEauthorblockA{\textit{dept. name of organization (of Aff.)} \\
% \textit{name of organization (of Aff.)}\\
% City, Country \\
% email address or ORCID}
% }

\maketitle

\vspace{-3mm}

\begin{abstract}
Multiple-input multiple-output (MIMO) transceiver and probabilistic shaping (PS) are key enablers for high spectral efficiency in 6G wireless networks. This work proposes a distribution-aware MIMO transceiver optimized for PS constellation symbols, including a Bayesian geometric mean decomposition (BGMD) precoder and a maximum a posteriori VBLAST (MAP-VBLAST) detector. The BGMD precoder uses PS priors in the derivation and equalizes layer gains to facilitate a single modulation and coding scheme for low-complexity transmissions while preserving channel capacity. MAP-VBLAST leverages these PS priors for optimal MAP detection within a successive interference cancellation (SIC) framework. Furthermore, a new codeword-to-layer mapping scheme, termed layer-contained MIMO (LC-MIMO), is proposed. By containing each codeblock (CB) within a single layer, LC-MIMO enables SIC at the CB level, allowing the receiver to exploit the error-correction capability of channel coding to mitigate error propagation. Numerical results show that the BGMD transceiver with LC-MIMO achieves notable performance gains over state-of-the-art methods.
\end{abstract}

\begin{IEEEkeywords}
GMD, UCD, transceiver, probabilistic shaping.
\end{IEEEkeywords}

% \nocite{brinton2025key, ivanov2025probabilistic,bocherer2015bandwidth, schulte2015constant, jiang2005joint, jiang2005ucd, yang2021joint, 3gpp38212, 3gpp38901, studer2008soft, hu2023supporting}
\vspace{-0.8em}
\section{Introduction}
% \vspace{-3mm}
The pursuit of higher spectral efficiency with reduced computational complexity has become a central objective in 6G wireless networks \cite{brinton2025key}.
Probabilistic shaping (PS) and multiple-input multiple-output (MIMO) transceiver design are key enablers for closing the gap to information-theoretic limits.
By tailoring the symbol distribution to approximate Gaussian inputs, PS enables communication systems to approach the Shannon channel capacity.
% PS tailors the modulation symbol distribution to better approximate Gaussian inputs for approaching the Shannon channel capacity.
This potential has motivated growing interest in integrating PS into modern wireless standards, including 5G New Radio (5G NR) \cite{hu2023supporting}, 6G, and Wi-Fi \cite{fang2025probabilistic}.

Conventional MIMO transceivers employ singular value decomposition (SVD)-based precoding to diagonalize the MIMO channel, but SVD could result in subchannels with unequal gains, which is undesirable for transmissions that employ a single modulation and coding scheme (MCS). 
In 5G NR, the codeword-to-layer (CW-to-layer) mapping distributes the bits of a codeblock (CB) across multiple spatial layers to exploit spatial diversity.
However, when the subchannel gains are highly unequal, this CW-to-layer mapping in 5G NR significantly degrades channel coding performance, particularly at high code rates.
To address this issue, geometric mean decomposition (GMD)-based transceivers \cite{jiang2005joint, jiang2005ucd} have been proposed to transform the MIMO channel into parallel subchannels with identical gains, improving the decoding performance for single-MCS transmission while preserving the Gaussian-input MIMO channel capacity.
Despite its structural compatibility with the transmission of PS quadrature amplitude modulation (QAM) symbols, existing GMD-based transceiver designs are derived under the assumption of uniformly distributed QAM symbols and cannot exploit the non-uniform symbol statistics induced by PS, leading to a performance gap relative to the achievable capacity under PS QAM signaling.

% the bits of a codeblock (CB) are distributed across multiple layers to exploit spatial diversity, while unequal subchannel gains can severely degrade channel coding and decoding performance, particularly in the high code-rate regime. 
% To address this issue, geometric mean decomposition (GMD)-based precoding \cite{jiang2005joint, jiang2005ucd} can be employed. 
% GMD decomposes the MIMO channel into parallel subchannels with identical gains, enabling reduced-complexity transmission using a single modulation and coding scheme while preserving Gaussian‑input MIMO channel capacity.

% allows the use of single modulation and coding scheme
% and preserves channel capacity with the use of a single modulation and coding scheme.

% , which decomposes the MIMO channel into parallel subchannels with identical gains and preserves capacity with single modulation and coding scheme can be used to resolve this problem.

% Despite the compatibility with PS QAM symbols, existing GMD-based transceiver designs cannot exploit the symbol distribution as a statistical prior and therefore exhibit a performance gap relative to the achievable capacity of MIMO channels.

Several recent works \cite{yang2021joint, kang2022probabilistic, hu2023supporting, ivanov2025probabilistic} have investigated PS in multiple-antenna systems.
The work \cite{yang2021joint} jointly optimizes PS constellation distributions and beamforming to maximize mutual information, but it is designed for multiple-input single-output systems. 
The work \cite{kang2022probabilistic} focuses on achieving shaping and diversity gains in MIMO channels through signal space diversity. 
The work \cite{hu2023supporting} investigates PS in 5G NR MIMO and demonstrates that PS can mitigate large signal-to-noise ratio (SNR) gaps between MCSs, improving the achievable throughput envelope.
The work \cite{ivanov2025probabilistic} studies PS symbols over MIMO channels using nonlinear sphere decoding \cite{studer2008soft}, showing shaping gains beyond the $1.53$ dB limit in additive white Gaussian noise (AWGN) channels. 
However, these works \cite{yang2021joint, kang2022probabilistic, hu2023supporting, ivanov2025probabilistic} do not consider a MIMO transceiver optimized for the PS symbol distribution.
% This letter focuses on developing a distribution‑aware MIMO transceiver that fully exploits probabilistic shaping gains under spatial multiplexing.
Further work is required to develop a distribution-aware MIMO transceiver capable of fully harvesting the PS gains in spatial multiplexing.

This work proposes a distribution-aware MIMO transceiver, Bayesian GMD (BGMD), that incorporates the PS symbol distribution as a statistical prior in the transceiver design.
\tcmajor{To the best of our knowledge, BGMD is the first GMD-based precoder that explicitly incorporates PS priors into the channel decomposition process and equalizes subchannel gains under PS signaling, co-designed with a MAP-VBLAST detector that exploits the non-uniform PS statistics for detection.}
% \tcmajor{A Bayesian GMD (BGMD) precoder is proposed as the first GMD-based linear precoder to embed PS priors into the channel decomposition process and equalize subchannel gains under PS signaling,
% % , embedding the shaping parameter into the augmented channel matrix to equalize subchannel gains under PS signaling, 
% co-designed with a MAP-VBLAST detector that exploits the non-uniform statistics of PS symbols for distribution-aware detection.}
To mitigate the susceptibility of SIC to error propagation, a layer‑contained MIMO (LC‑MIMO) mapping scheme is proposed. 
LC-MIMO assigns each CB to a single spatial layer, rather than distributing it across layers as in 5G NR, making CB-level SIC feasible.
This allows the receiver to exploit the error‑correction capability of channel coding, producing more reliable symbol estimates that mitigate error propagation in SIC.
Numerical results show that the BGMD transceiver with CB-SIC achieves substantial gains in both throughput and decoding performance compared with state‑of‑the‑art methods.

{\bf Notation:}	Bold lowercase letters $\mathbf{x}$ and bold uppercase letters $\mathbf{X}$ denote vectors and matrices, respectively;	$\mathbf{X}^\top$, $\mathbf{X}^H$, $\mathbf{X}^{-1}$ denote the transpose, conjugate transpose, and inverse of $\mathbf{X}$, respectively. $\mathbf{I}_N$ denotes the $N\times N$ identity matrix.
    % ; $[\mathbf{X}]_{m,n}$ represents the $(m,n)$-th element of $\mathbf{X}$. %; $\lfloor x \rceil$ denotes the nearest integer to $x$.

\section{System Model}
Section \ref{subsection_signal_model} introduces the signal model with linear MIMO precoding. Section \ref{Sec_PS_Symbol} describes the probabilistic shaping method used for QAM symbol generation.

\subsection{Signal Model}\label{subsection_signal_model}
We consider a single-user MIMO downlink transmission with $N_t$ transmit antennas and $N_r$ receive antennas, employing spatial multiplexing with $L\leq \min(N_t,N_r)$ spatial layers.
% from an $N_t$-antenna base station to an $N_r$-antenna user.
% Spatial multiplexing is employed with $L\leq \min(N_t,N_r)$ transmission layers.
% Let $L\leq \min(N_t,N_r)$ denote the number of layers for spatial multiplexing, 
The received signal $\mathbf{y}\in\mathbb{C}^{N_r \times 1}$ is given by
\begin{equation}\label{eq_signal_model}
    \mathbf{y} = \mathbf{H}\mathbf{F}\mathbf{x} + \mathbf{n},
\end{equation}
where $\mathbf{H}\in\mathbb{C}^{N_r \times N_t}$ is the MIMO channel, $\mathbf{F}\in\mathbb{C}^{N_t\times L}$ is a linear precoder, and $\mathbf{x}\in\mathbb{C}^{L \times 1}$ is the transmitted symbol vector.
The noise $\mathbf{n}\sim\mathcal{CN}(\mathbf{0},\sigma_n^2\mathbf{I}_{N_r})$ is additive white Gaussian noise.
% and $\mathbf{x}=[x_1,x_2,\dots,x_L]^{\top}\in\mathbb{C}^{L\times 1}$ is the symbol vector across the $L$ spatial layers. 
% The additive noise $\mathbf{n}$ is an independent and identically distributed (i.i.d.) complex Gaussian random variable, i.e., $\mathbf{n}\sim\mathcal{CN}(\mathbf{0},\sigma_n^2 \mathbf{I}_{N_r})$.
Although a flat‑fading channel is assumed, this work can be extended to frequency‑selective channels via orthogonal frequency‑division multiplexing, where each subcarrier experiences an equivalent flat‑fading channel.
\tcmajor{The proposed precoder can be extended to multi-user MIMO systems, with either time-domain or frequency-domain multi-user access methods (TDMA or FDMA), or on top of the leakage-based multi-user precoding schemes~\cite{sadek2007leakage, cheng2010new}.}

\subsection{Probabilistic Shaping of QAM Symbols}\label{Sec_PS_Symbol}
% Leveraging the fact that Gaussian signaling achieves capacity over the AWGN channel, 
Leveraging the capacity-achieving property of Gaussian signaling over the AWGN channel,
we employ QAM symbols shaped by the PS method\footnote{Also known as probabilistic amplitude shaping as in \cite{bocherer2015bandwidth} or probabilistic constellation shaping as in \cite{hu2023supporting}.}~\cite{bocherer2015bandwidth}, denoted as PS QAM symbols.
A $2^{2M}$‑QAM symbol is constructed as the Cartesian product of two $2^{M}$‑PAM constellations, each with alphabet $\mathcal{A}=\{\pm1,\pm3,\dots,\pm(2^{M}-1)\}$.
Following \cite{bocherer2015bandwidth}, PS is applied to the PAM amplitudes, and the sign bits are assumed to be uniformly distributed. A constant composition distribution matcher (CCDM) \cite{schulte2015constant} is adopted to map uniformly distributed data bits to shaped amplitude sequences according to a Maxwell–Boltzmann (MB) distribution\footnote{\tcmajor{An example of a PS 64-QAM constellation with MB distribution is shown in Fig.~1 of \cite{hu2023supporting}.}}
% \footnote{\tcmajor{Any probabilistic shaping scheme capable of producing symbols following the target MB distribution may be employed for symbol construction.}}
\begin{equation}\label{eq_Maxwell_Boltzmann_distribution}
    \mathbb{P}_{\nu}(s)=\frac{e^{-\nu\lVert s\rVert^2}}{\zeta},\quad s\in\mathcal{A},
\end{equation}
where $\nu$ is the shaping parameter and $\zeta=\sum_{s^\prime\in\mathcal{A}} e^{-\nu\lVert s^\prime\rVert^2}$ is the normalization constant. 
% \begin{remark}
\tcmajor{Note that, in addition to CCDM, any distribution matcher capable of generating symbols with the target MB distribution, e.g., enumerative sphere shaping \cite{gultekin2019enumerative}, can be compatible with our proposed MIMO transceiver.}
% \tcmajor{Note that this work focuses on MIMO transceiver design for PS QAM symbols, where  we consider the CCDM as the distribution matcher. Any distribution matcher capable of generating symbols with the target MB distribution, e.g., enumerative sphere shaping \cite{gultekin2019enumerative}, can be employed with our transceiver design.}
% \end{remark}
% \tcmajor{Note that this work focuses on the MIMO transceiver design for the PS symbols following this MB distributions, where any PS schemes, e.g., Enumerative Sphere shaping \cite{gultekin2019enumerative}, capable of generating symbols  following the target MB distribution may be emplyed.
% Note that this work is not confined to CCDM for PS symbol construction. Any PS schemes, e.g., , capable of generating symbols following the target MB distribution may be emplyed.
% }

To satisfy the average transmit power constraint, the transmitted symbol vector is scaled as
\begin{equation}\label{eq_PS_power_scaling}
    \mathbf{x} = \alpha \mathbf{s},
\end{equation}
where $\mathbf{s}=[s_1,\dots,s_L]^{\top}$ is the symbol vector across the $L$ spatial layers. The scaling factor $\alpha>0$ is chosen such that $\mathbb{E}[|\alpha  s_\ell|^2]=P_t/L$, where $P_t$ is the total transmit power.

\section{Distribution-Aware GMD Transceiver}
Here, we propose a distribution-aware GMD transceiver for PS QAM symbols.
First, we design a MAP detector with a VBLAST structure (MAP-VBLAST) in Section \ref{subsection_MAP_estimator} and a BGMD precoder exploiting the PS priors in Section \ref{subsection_BGMD}.
Then, we introduce the LC-MIMO mapping to enable SIC at the CB level for mitigating error propagation in Section \ref{subsection_LCmimo}.

% Section \ref{subsection_MAP_estimator} develops the optimal MAP estimator considering the PS prior, applied within a VBLAST-like layer-by-layer symbol detection.
% Section \ref{subsection_BGMD} proposes the .
% Section \ref{subsection_LCmimo} introduces the LC-MIMO mapping.

% With the priors of the symbols induced by PS symbols, we derive the MAP estimator for the symbol detector in MIMO systems with a given linear precoder.
% Based on that, we 

% Considering the VBLAST symbol detection, 
% % We first derive the MAP-based Bayesian estimator for the PS symbols.
% After the MAP derivation, we point out that the layer imbalance issue, and proposed the Bayesian GMD precoding.
% Since the Bayesian GMD precoding applies VBLAST architecture, in which the error propagation may greatly degrades the decoding performance. To address the issue, we propose a novel LC-MIMO mapping, which reduces the chance of error propagation in VBLAST structure by channel coding.

% \vspace{-1mm}
\subsection{MAP-VBLAST Symbol Detection}\label{subsection_MAP_estimator}
We derive the MAP detection \cite{kay1993fundamentals} for the PS QAM symbols, defined as
\begin{align}
  \mathbf{\hat s}_{map}= \arg\max_{\mathbf{s}}\left\{\ln \mathbb{P}(\mathbf{y}|\mathbf{s}) + \ln \mathbb{P}(\mathbf{s})\right\},  
\end{align}
where $\mathbb{P}(\mathbf{y}|\mathbf{s})$ is the conditional probability of the received signal given the transmitted symbol, and $ \mathbb{P}(\mathbf{s}) $ is the statistical prior of the transmitted symbols.
Substituting the transmitted symbol vector in \eqref{eq_PS_power_scaling} into the signal model in \eqref{eq_signal_model}, we have $\mathbf{y}=\alpha\mathbf{H}\mathbf{F}\mathbf{s}+\mathbf{n}$.
With PS, the transmitted symbols follow the MB distribution  $\mathbb{P}_{\nu}(\mathbf{s})$ in \eqref{eq_Maxwell_Boltzmann_distribution}, which serves as the statistical prior (or PS priors), yielding $\ln \mathbb{P}(\mathbf{s})=\ln \mathbb{P}_{\nu}(\mathbf{s}) = -\nu\lVert\mathbf{s}\rVert^2 - \ln{\zeta}$.
% The PS technique employs the MB distribution $\mathbb{P}_{\nu}(\mathbf{s})$ in \eqref{eq_Maxwell_Boltzmann_distribution} to shape the symbol constellation, leading to the statistical prior (or PS priors) of the transmitted symbols. 
% Thus, we have $\ln \mathbb{P}(\mathbf{s})=\ln \mathbb{P}_{\nu}(\mathbf{s}) = -\nu\lVert\mathbf{s}\rVert^2 - \ln{\zeta}$.
Therefore, the MAP detection can be reformulated as
\begin{align}
    \mathbf{\hat s}_{map}&=\arg\min_{\mathbf{s}} \left\{\frac{\lVert\mathbf{y}-\alpha\mathbf{H}\mathbf{F}\mathbf{s}\rVert^2}{\sigma_n^2}+\nu\lVert\mathbf{s}\rVert^2\right\}\\
    \label{eq_map_estimation_aug_channel}
    &=\arg\min_{\mathbf{s}} \left\{\left\lVert
    \begin{bmatrix}
        \mathbf{y}\\
        0
    \end{bmatrix}
    -
    \begin{bmatrix}
        \alpha\mathbf{H}\mathbf{F}\\
        \sqrt{\sigma_n^2\nu}\mathbf{I}_L
    \end{bmatrix} \mathbf{s}
    \right\rVert^2
    \right\}.
\end{align}
% where $\mathbf{I}$ denotes the identity
Denoting the augmented channel matrix
\begin{align}\label{eq_aug_channel_V1}
    \mathbf{G} &= \begin{bmatrix}
        \alpha\mathbf{H}\mathbf{F}\\
        \sqrt{\sigma_n^2\nu}\mathbf{I}_L
    \end{bmatrix},
\end{align}
we perform a QR decomposition of $\mathbf{G}$, given by
\begin{align} %\nonumber
    \mathbf{G} = \mathbf{Q}_{G}\mathbf{R}_{G} \triangleq
  \begin{bmatrix}
      \mathbf{Q}_{G}^{\mathcal{U}}\\
      \mathbf{Q}_{G}^{\mathcal{L}}
  \end{bmatrix}\mathbf{R}_G,
  \label{eq_aug_channel}
\end{align}
where $\mathbf{Q}_{G}$ is an $(N_r + L)\times L$ matrix with orthonormal columns, $\mathbf{Q}_{G}^{\mathcal{U}}$ is an $N_r\times L$ matrix, and $\mathbf{R}_{G}$ is an $L\times L$ upper triangular matrix with positive diagonal elements. 
The MAP detection of $\mathbf{s}$ can be reformulated by multiplying $({\mathbf{Q}_G})^H$ on the objective function in \eqref{eq_map_estimation_aug_channel}, given by
\begin{align} \label{eq_map_estimate_afterQR}
    \mathbf{\hat{s}}_{map}=\arg\min_{\mathbf{s}} \left\{\left\lVert
    \mathbf{\tilde y}    
    -
    \mathbf{R}_{G} \mathbf{s}
    \right\rVert^2
    \right\}=\mathbf{R}_{G}^{-1}\mathbf{\tilde y},
\end{align}
where $\mathbf{\tilde y}=[\tilde y_1,\dots,\tilde y_L]^{\top} = \left(\mathbf{Q}_{G}^{\mathcal{U}}\right)^H \mathbf{y}$.
With the upper-triangular matrix $\mathbf{R}_{G}$, we operate the VBLAST-like procedure involving sequential nulling and cancellation, with the detection order from layer $L$ to layer $1$, detailed as
\begin{align}
    {\tilde{s}}_{map,i} = \mathcal{M}\left(\frac{\tilde{y}_i - \sum_{j= i +1}^{L}[\mathbf{R}_{G}]_{i,j}{\tilde{s}}_{map,j}}{[\mathbf{R}_{G}]_{i,i}}\right),
\end{align}
from $i=L$ to $i=1$, where ${\tilde{s}}_{map,i}$ denotes the $i$-th element of ${\tilde{\mathbf s}}_{map}$ and  $[\mathbf{R}_{G}]_{i,j}$ represents the $(i,j)$-th element of $\mathbf{R}_{G}$. 
$\mathcal{M}(\cdot)$ denotes the symbol mapping in the QAM constellation.

\subsection{Bayesian GMD Precoder} \label{subsection_BGMD}
The primary challenge in VBLAST detection is its susceptibility to error propagation arising from SIC. 
Conventional approaches \cite{hassibi2000efficient} mitigate this issue by optimizing the detection order based on post‑detection SNR, at the cost of increased receiver complexity.
In this subsection, we propose a BGMD precoder that equalizes the effective layer gains across layers, eliminating the need for optimal detection ordering. The resulting equal layer gains of BGMD also enable the use of a single-MCS scheme for low-complexity transmissions while preserving channel capacity. Furthermore, by embedding PS priors into the precoder design, BGMD can provide enhanced throughput and decoding performance compared to conventional GMD‑based designs \cite{jiang2005joint, jiang2005ucd}. %, as shown in Fig. \ref{fig:bler_comparison} and Fig. \ref{fig:capacity_comparison}. 

% With the MAP detector in \eqref{eq_map_estimate_afterQR}, the precoder $\mathbf{F}$ is designed such that the diagonal elements of $\mathbf{R}_G$ are equal, yielding uniform decoding SNR across all layers. This enables the use of a single modulation and coding scheme (MCS) while preserving channel capacity.

With the MAP detector in \eqref{eq_map_estimate_afterQR}, we seek to design the BGMD precoder $\mathbf{F}$ such that the diagonal elements of $\mathbf{R}_{G}$ are equal.
The SVD of the MIMO channel is given by $\mathbf{H}=\mathbf{U}\mathbf{S}\mathbf{V}^H$, where $\mathbf{U}\in\mathbb{C}^{N_r\times L}$ and $\mathbf{V}\in\mathbb{C}^{N_t\times L}$ are semi-unitary matrices, and $\mathbf{S}\in\mathbb{R}^{L\times L}$ is a diagonal matrix with the singular values.
We assume the BGMD precoder takes the form
\begin{align}\label{eq_bayesian_GMD}
    \mathbf{F} = \mathbf{V}\mathbf{\Phi}^{1/2}\mathbf{\Omega}^H,
\end{align}
where $\mathbf{\Phi}=\mathrm{Diag}\{\phi_1, \dots,\phi_L\}$ is the power-loading matrix, and $\mathbf{\Omega}$ is a unitary matrix to be designed for layer mixing. 
Substituting $\mathbf{F}$ (as in \eqref{eq_bayesian_GMD}) into the augmented MIMO channel $\mathbf{G}$ (as in \eqref{eq_aug_channel_V1}), we have 
\begin{align}\label{eq_aug_channel_V2}
    \mathbf{G} &= \begin{bmatrix}
        \alpha\mathbf{U}\mathbf{S}\mathbf{\Phi}^{1/2}\mathbf{\Omega}^H\\
        \sqrt{\sigma_n^2 \nu} \mathbf{I}_L
    \end{bmatrix}\\
    &=\begin{bmatrix}
        \mathbf{I}_{N_r} & \mathbf{0}\\
        \mathbf{0} & \mathbf{\Omega}
    \end{bmatrix}
    \begin{bmatrix}
        \alpha\mathbf{U}\mathbf{S}\mathbf{\Phi}^{1/2}\\
        \sqrt{\sigma_n^2 \nu} \mathbf{I}_L
    \end{bmatrix}\mathbf{\Omega}^H.
    \label{eq_aug_channel_V3}
\end{align}
% where the second equality follows from the unitary property of $\mathbf \Omega$.
The second equation holds because $\mathbf{\Omega}$ is a unitary matrix. 
We apply the GMD \cite{jiang2005joint} to the middle matrix in \eqref{eq_aug_channel_V3} as 
\begin{align}\label{eq_GMD}
    \mathbf{B}\triangleq
    \begin{bmatrix}
        \alpha\mathbf{U}\mathbf{S}\mathbf{\Phi}^{1/2}\\
        \sqrt{\sigma_n^2 \nu} \mathbf{I}_L
    \end{bmatrix} = \mathbf{Q}_{B}\mathbf{R}_B \mathbf{P}_B^H,
\end{align}
where $\mathbf{Q}_{B}$ is a semi-unitary matrix, $\mathbf{P}_B$ is a unitary matrix, and $\mathbf{R}_B$ is an upper triangular matrix whose diagonal elements are identical and equal to the geometric mean of the singular values of $\mathbf{B}$.
% with equal diagonals having the value of the geometric mean of the SVD values of $\mathbf B$.
By choosing $\mathbf{\Omega}^H=\mathbf{P}_B$ in \eqref{eq_aug_channel_V3}, the augmented channel can be expressed as $\mathbf{G}=\tilde{\mathbf{Q}}\mathbf{R}_B$, where
\begin{equation}
    \tilde{\mathbf{Q}}=
    \begin{bmatrix}
        \mathbf{I}_{N_r} & \mathbf{0}\\
        \mathbf{0} & \mathbf{\Omega}
    \end{bmatrix}
    \mathbf{Q}_B.
\end{equation}
Accordingly, the BGMD precoder is obtained by substituting $\mathbf{\Omega}^H=\mathbf{P}_B$ into \eqref{eq_bayesian_GMD}, yielding $\mathbf{F}=\mathbf{V}\mathbf{\Phi}^{1/2}\mathbf{P}_B$.
This BGMD precoder $\mathbf{F}$ consists of the matrix $\mathbf{V}$ for channel orthogonalization, the matrix $\mathbf{\Phi}$ for power‑loading, and the unitary matrix $\mathbf{P}_B$ for equalizing the effective subchannel gains. 
Since $\mathbf{P}_B$ depends on the shaping parameter $\nu$ through \eqref{eq_GMD}, the BGMD precoder is explicitly distribution‑aware.

\tcmajor{Compared with the uniform channel decomposition (UCD) transceiver~\cite{jiang2005ucd}, the proposed BGMD precoder is better matched to MAP detection for PS symbols by explicitly incorporating the shaping parameter $\nu$ into the transceiver design.
UCD is originally designed with the MMSE-VBLAST, which treats uniform QAM symbols as Gaussian with variance $\sigma_s^2$ (average symbol power), effectively fixing $\nu = 1/\sigma_s^2$ in \eqref{eq_aug_channel_V1}.
Under PS, however, the true shaping parameter generally satisfies $\nu \neq 1/\sigma_s^2$, rendering UCD mismatched to the actual symbol  distribution. In contrast, BGMD jointly exploits the channel and the true shaping parameter, ensuring consistency with the MAP detection. As demonstrated in Section~\ref{sec:numerical_results}, this distribution-aware BGMD precoder design yields improved throughput and decoding performance compared with UCD.}

\subsection{Layer-Contained MIMO (LC-MIMO) and CB-SIC}\label{subsection_LCmimo}
Although the BGMD precoder equalizes the post‑detection SNR across spatial layers and eliminates the need for optimal detection ordering in SIC, symbol detection errors may still occur during SIC.
% In 5G NR, CW‑to‑layer mapping follows a standardized NR‑MIMO procedure \cite{3gpp38211} that distributes the coded bits (of a CB) across multiple spatial layers.
Symbol detection errors can propagate across layers during SIC and degrade decoding performance, potentially increasing overhead in the hybrid automatic repeat request process and retransmissions for reliable recovery.

To address this issue, LC‑MIMO performs CW‑to‑layer mapping at the CB granularity, assigning each CB to a single spatial layer, rather than following the standardized 5G NR procedure that distributes coded bits across multiple layers \cite{3gpp38211}.
Each CB is confined to a single spatial layer, enabling SIC to operate at the CB level instead of the coded‑bit level.
This allows the receiver to exploit the error‑correction capability of channel coding to obtain more reliable symbol estimates and reduce error propagation in SIC.

% Such errors can become a performance bottleneck, necessitating additional overhead in the hybrid automatic repeat request (HARQ) process and retransmissions for reliable recovery. 
% Therefore, we propose a novel CW-to-layer mapping scheme, termed LC-MIMO, to enable CB-level SIC (CB-SIC) and mitigate cross-layer error propagation in MAP-VBLAST.

% In 5G NR, the CW-to-layer mapping follows a standardized NR-MIMO procedure \cite{3gpp38211}, in which the coded bits (of a CB) are distributed across multiple layers to exploit spatial layer diversity. However, symbol detection errors may still occur and propagate across layers during SIC, leading to severe degradation in decoding performance.
% To address this issue, LC-MIMO performs CW-to-layer mapping at the granularity of CBs rather than coded bits in NR‑MIMO.
% Each CB is confined to a single spatial layer, enabling SIC to operate at the CB level instead of the coded‑bit level. 
% This allows the receiver to exploit the error‑correction capability of channel coding to obtain more reliable symbol estimates and reduce error propagation in SIC.

\begin{figure}[t]		\centering %\setcounter{figure}{0}
		\includegraphics[width=\linewidth]{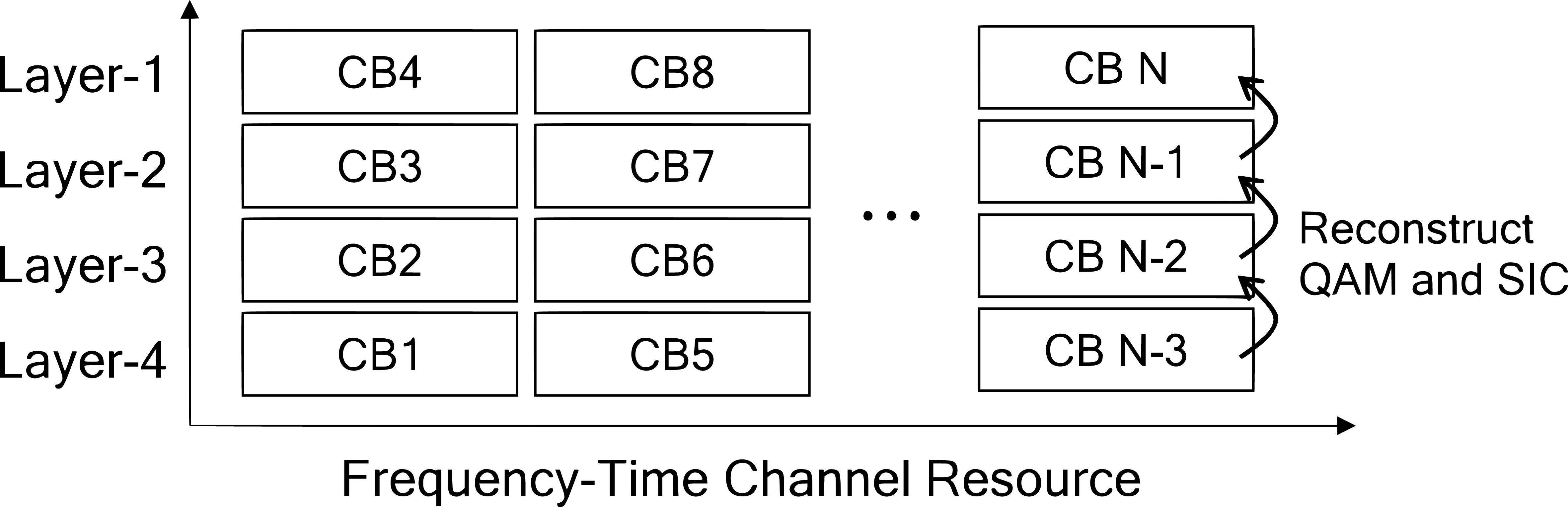}
		\caption{
			LC-MIMO and the illustration of CB-SIC.
		}
		\label{fig:LC_mimo_and_CBsic}
		\vspace{-3mm}
\end{figure}

The CB-SIC procedure operates as follows. 
% Assuming a CW consisting of $N$ CBs as in Fig.~\ref{fig:LC_mimo_and_CBsic}, CB-SIC can be operated on the CBs occupying the same channel resource, e.g., CB $N-3$ ~ CB $N$.
First, the symbols corresponding to a CB on a specific layer are estimated and demapped into coded bits. These bits are then processed by a channel decoder (e.g., an LDPC decoder) to correct bit errors stemming from symbol detection errors.
Subsequently, the decoded bits are re-encoded (e.g., by an LDPC encoder) and modulated to reconstruct the QAM symbols, providing significantly more reliable estimates.
These reconstructed symbols are then subtracted from the received signal to cancel inter-layer interference. 
By performing interference cancellation using decoded and re-encoded symbols, rather than unreliable hard-symbol estimates, CB-SIC substantially reduces the likelihood of error propagation.
% , which facilitates the decoding for the remaining CB $N-2$ ~ CB $N$. 
LC-MIMO and the CB-SIC are illustrated in Fig.~\ref{fig:LC_mimo_and_CBsic}.

\section{Complexity Analysis}
\tcmajor{
% BGMD is dominated by the SVD with a complexity of $\mathcal{O}(N_r N_t L)$. 
% MAP-VBLAST relies on a QR decomposition, with a complexity of $\mathcal{O}((N_r+L)L^2)$, and the SIC involves triangular back-substitution with a complexity of $\mathcal{O}(L^2)$. 

BGMD is dominated by the SVD with a complexity of $\mathcal{O}(N_r N_t L)$.
MAP-VBLAST relies on QR decomposition with $\mathcal{O}((N_r+L)L^2)$, 
and SIC via back-substitution with $\mathcal{O}(L^2)$.

CB-SIC comprises LDPC decoding, re-encoding, and symbol mapping. 
LDPC decoding is the dominant cost at $\mathcal{O}(n_{\text{bit}} \cdot N_{\text{itr}})$, where $n_{\text{bit}}$ is the number of coded bits per CB and $N_{\text{itr}}$ is the number of decoding iterations (typically 20--50). 
LDPC re-encoding and symbol mapping are both with the complexity of $\mathcal{O}(n_{\text{bit}})$. 
Since LDPC decoding is standard receiver processing, the additional 
CB-SIC overhead, including re-encoding and symbol mapping, is modest. In terms of latency, LC-MIMO enables CB-SIC to operate on groups of $L$ CBs sharing the same channel resource, so only $L$ CBs need to be buffered per processing group, which keeps the latency limited.

For PS symbols, CCDM maps uniformly distributed input bits 
to a sequence of $n_c$ amplitude symbols drawn from alphabet $\mathcal{A}$. 
Using an arithmetic coding-based implementation~\cite{schulte2015constant}, 
the matcher and dematcher each scale linearly as 
$\mathcal{O}(n_c \cdot |\mathcal{A}|)$.

}

\section{Numerical Results}\label{sec:numerical_results}
% \subsection{Experiment Setting}
In this section, we evaluate the performance of the BGMD precoder and MAP-VBLAST, integrated with LC-MIMO mapping and CB-SIC, for the transmission of PS QAM symbols.
We consider a single-user downlink MIMO transmission having $N_r=4$ receiver antennas, $N_t=4$ transmit antennas, and $L=4$ spatial layers.
We assume the $20$\,MHz TDL-A MIMO channel in 5G NR \cite{3gpp38901}, corresponding to $48$ resource blocks, each comprising $12$ subcarriers at $30$\,kHz subcarrier spacing.
The channel exhibits a $30$\,ns delay spread and an $11$\,Hz Doppler spread.
The 5G NR low-density parity-check (LDPC) code \cite{3gpp38212} and the PS QAM symbol generation \cite{bocherer2015bandwidth} are employed. 
The input data bits are generated by CCDM \cite{schulte2015constant}, which follows an MB distribution with a shaping parameter $\nu$, as defined in \eqref{eq_Maxwell_Boltzmann_distribution}.
We consider $64$-QAM symbols shaped by PS with a shaping parameter $\nu$ and an LDPC code rate of $0.9258$ for evaluation.
Equal power allocation across the $L$ layers is assumed.
We assume that each transport block (TB) is transmitted within a single slot of duration $T_{\text{slot}} = 0.5$\,ms and corresponds to a single CW.
This CW is segmented into $N_{\text{cb}}$ LDPC CBs, where each CB contains $N_{\text{cbit}}$ coded bits determined by the 5G NR rate‑matching procedure \cite{3gpp38212}.

We consider the following baseline schemes for comparison. For transceiver design, the UCD transceiver~\cite{jiang2005ucd} is adopted as the baseline. This UCD transceiver represents the state‑of‑the‑art GMD‑based precoding with MMSE‑VBLAST while not exploiting the PS priors.
{For symbol detection, sphere decoding (SD) \cite{studer2008soft} is included as a benchmark. SD provides near maximum‑likelihood detection performance but incurs higher computational complexity than linear detectors such as MAP or MMSE.}
Since LC‑MIMO enables CB‑SIC, we include a Hard‑SIC scheme under NR‑MIMO mapping as a benchmark to highlight the impact of error propagation, where Hard‑SIC performs interference cancellation based on hard‑symbol estimates.
% Since LC-MIMO enables the CB-SIC, to highlight the impact of error propagation, we include "Hard-SIC" in NR-MIMO mapping as a benchmark, where Hard-SIC is performed on hard-sliced symbols. 
For CW‑to‑layer mapping, the mapping specified in 5G NR is used as a baseline, denoted as NR‑MIMO \cite{3gpp38211}.

\begin{figure}[t]
    \centering
    \begin{subfigure}[b]{0.48\textwidth}
        \centering
        \includegraphics[scale=0.5]{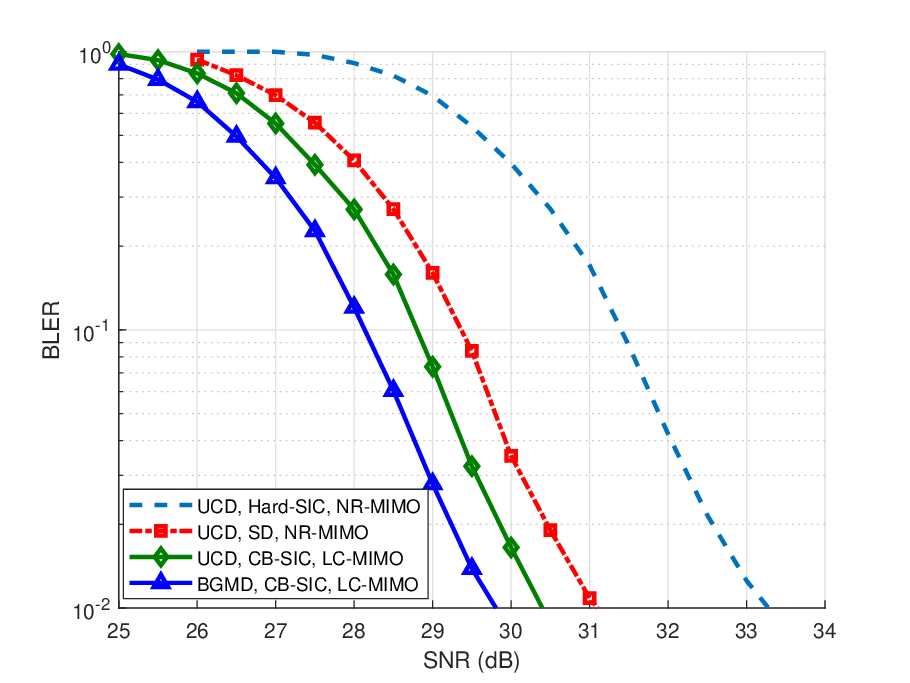}
        \caption{Shaping parameter $\nu = 0.05$.}
        \label{fig:bler_v0p05}
    \end{subfigure}
    \hfill
    \begin{subfigure}[b]{0.48\textwidth}
        \centering
        \includegraphics[scale=0.5]{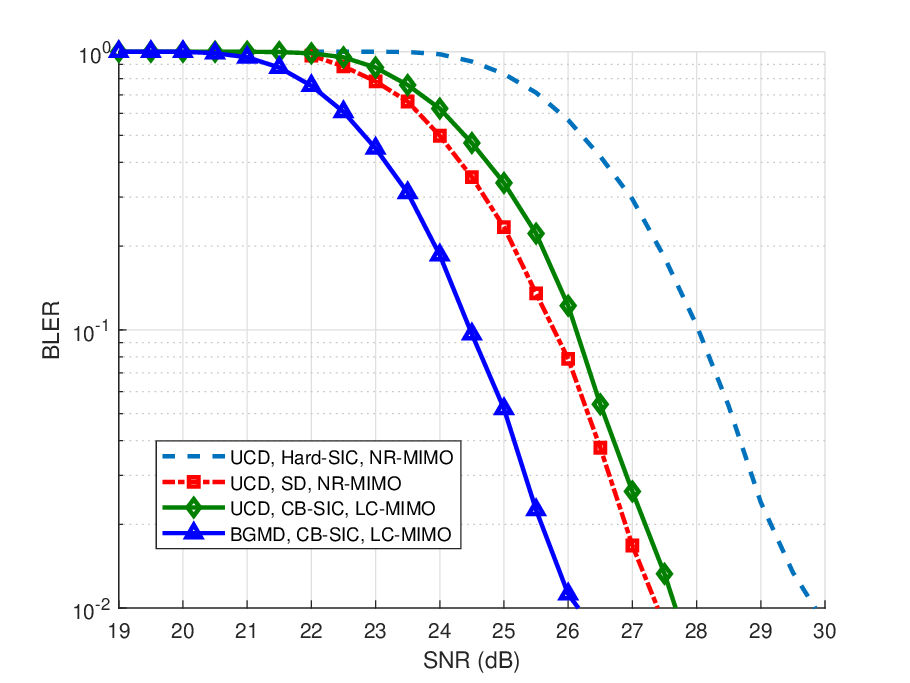}
        \caption{Shaping parameter $\nu = 0.1$.}
        \label{fig:bler_v0p1}
    \end{subfigure}
    \caption{BLER versus SNR.}% \nt{Remove the line for UCD with MAP}}
    \label{fig:bler_comparison}
    \vspace{-3.5mm}
\end{figure}

% \subsection{Decoding Performance}
Fig. \ref{fig:bler_comparison} presents the block error rate (BLER) versus SNR for various transceiver schemes and shaping parameters. 
The BLER corresponds to the CW error rate, where a CW is considered erroneous if any CB within the CW fails the cyclic redundancy check.
\tcmajor{In Fig.~\ref{fig:bler_v0p05} ($\nu = 0.05$), the UCD with CB-SIC achieves 10\% BLER at $28.8$~dB, providing a $0.5$~dB gain over UCD with SD, confirming that CB-SIC achieves better decoding performance at lower complexity than the nonlinear SD under UCD precoding. The UCD with Hard-SIC requires $31.2$~dB, a $2.4$~dB degradation relative to CB-SIC, validating that CB-SIC effectively mitigates error propagation. BGMD with CB-SIC further reduces the required SNR to $28.1$~dB, yielding a $0.7$~dB gain over UCD with CB-SIC from joint distribution-aware precoding and symbol detection.
Fig.~\ref{fig:bler_v0p1} shows the BLER versus SNR for a larger shaping parameter ($\nu = 0.1$). The performance trends are consistent with those of $\nu = 0.05$, but the gains from the distribution-aware design are more pronounced, as the PS distribution departs more significantly from the uniform case. UCD with CB-SIC achieves a 10\% BLER at $26$~dB, outperforming its Hard-SIC counterpart by $2$~dB. Since UCD is designed for uniform QAM, it incurs a mismatch with the MAP detector under PS transmission with a larger $\nu = 0.1$, resulting in worse BLER performance than the UCD with SD. Once the PS prior is incorporated in the precoder design, BGMD with CB-SIC achieves 10\% BLER at only $24.5$~dB, outperforming UCD with CB-SIC ($26$~dB) by $1.5$~dB. This also shows a $1.3$~dB gain over the UCD with SD, demonstrating that BGMD with CB-SIC improves decoding reliability while significantly reducing demapper complexity. These results indicate that the BLER gain of BGMD scales with the shaping parameter (i.e., the degree of PS shaping).}

To compare the transceiver performance, we evaluate the empirical throughput versus SNR under different shaping parameters.
The empirical throughput\footnote{We impose the constraint of a fixed modulation order of 64 QAM.} is defined as 
\begin{align}
    R_{QAM} = \frac{1}{T_{slot}}\left(N_{cb}H(\mathbf{x}) - \sum_{i=1}^{N_{cb}}\sum_{j=1}^{N_{cbit}}H_{b}\left(p_j^{(i)}\right)\right),
\end{align}
where $H(\mathbf{x})$ denotes the source entropy of the PS QAM symbols within one CB. 
%$H_b(p)$ is the binary entropy function \cite{cover1999elements}.
$H_b(p) = -p\log_2p-(1-p)\log_2(1-p)$ is the binary entropy function.
$p_j^{(i)}$ represents the posterior probability of the $j$-th coded bit in the $i$-th CB, obtained from its log-likelihood ratio (LLR) $\ell_j^{(i)}$ at the output of the equalizer, i.e., $p_j^{(i)} = \left(1+e^{-\ell_j^{(i)}}\right)^{-1}$.
With the PS symbol structure \cite{bocherer2015bandwidth}, we have the source entropy per CB as $H(\mathbf{x}) = H(\mathbf{x}_{amp}) + H(\mathbf{x}_{sign})$, where $H(\mathbf{x}_{amp})$ and $H(\mathbf{x}_{sign})$ are the amplitude entropy and the sign entropy, respectively.
% PS imposes a structural decomposition of each transmitted symbol $\mathbf{x}=(\mathbf x_{amp},\mathbf x_{sign})$, where $\mathbf x_{amp}$ denotes the shaped amplitude bits and $\mathbf x_{sign}$ denotes the uniformly distributed sign bits. Hence, the source entropy per codeblock is $H(\mathbf{x}) = H(\mathbf{x}_{amp}) + H(\mathbf{x}_{sign}).$
The amplitude entropy is given by $H(\mathbf{x}_{amp})=\frac{M_{amp}}{\log_2(Q)/2-1}H_\nu$, where $M_{amp}$ is the number of amplitude bits per CB, $Q = 64$ for the 64-QAM modulation, and $H_\nu=-\sum_{s\in\mathcal{A}}\mathbb{P}_{\nu}(s)\log_2{\mathbb{P}_{\nu}(s)}$ is the MB shaping entropy per amplitude symbol under the shaping parameter $\nu$.
Since the sign bits are assumed to be uniformly distributed and therefore each carries one bit of entropy, the sign entropy per CB is $H(x_{sign}) = N_{cbit}- M_{amp}$.

\begin{figure}[t]
    \centering
    \begin{subfigure}[b]{0.48\textwidth}
        \centering
        \includegraphics[scale=0.5]{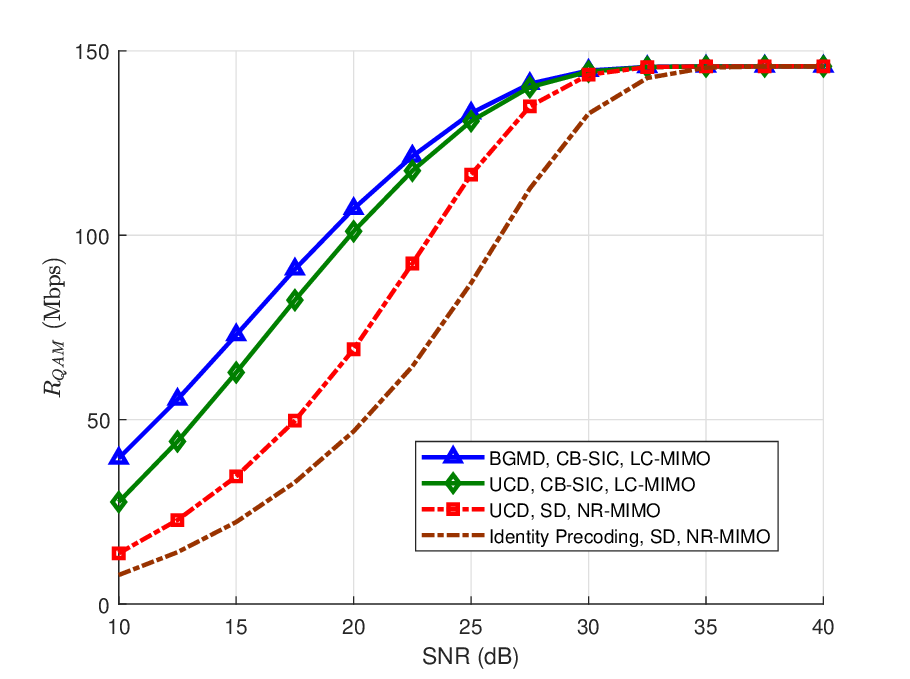}
        \caption{Shaping parameter $\nu = 0.05$.}
        \label{fig:capacity_v0p05}
    \end{subfigure}
    \hfill
    \begin{subfigure}[b]{0.48\textwidth}
        \centering
        \includegraphics[scale=0.5]{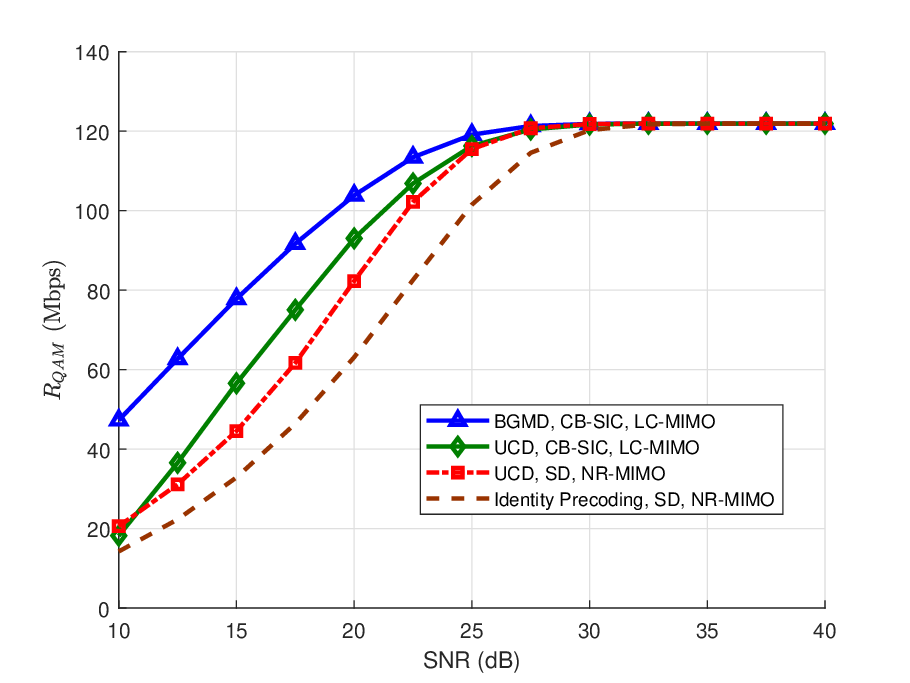}
        \caption{Shaping parameter $\nu = 0.1$.}
        \label{fig:capacity_v0p1}
    \end{subfigure}
    \caption{Empirical throughput versus SNR.}% \nt{Remove the line for UCD with MAP}}
    \label{fig:capacity_comparison}
    \vspace{-3mm}
\end{figure}

Fig. \ref{fig:capacity_comparison} illustrates the empirical throughput versus SNR for $\nu=0.05$ and $\nu=0.1$, where BGMD is benchmarked against the state-of-the-art UCD transceiver and identity precoding baselines.
% We evaluate BGMD performance against the state‑of‑the‑art UCD transceiver and identity precoding baselines.
% To evaluate BGMD, we compare its performance against the state‑of‑the‑art UCD transceiver and identity precoding baselines.
As shown in Fig. \ref{fig:capacity_v0p05} ($\nu=0.05$), identity precoding with SD requires $\text{SNR}=26$ dB to achieve a throughput of $100$ Mbps. 
Employing UCD with SD yields a $2$ dB SNR gain. 
% With the BGMD precoder, the proposed CB‑SIC in LC‑MIMO further improves performance, achieving a $4$ dB SNR improvement over the SD demapper in NR‑MIMO at the same throughput of $100$ Mbps. 
With UCD precoding, CB‑SIC outperforms SD by $3$ dB at $100$ Mbps, highlighting the benefits of inter‑layer interference cancellation via CB-SIC. 
Notably, BGMD with CB‑SIC consistently outperforms UCD with CB‑SIC, demonstrating the advantage of incorporating PS priors in the transceiver design.
Fig. \ref{fig:capacity_v0p1} shows results for a larger shaping parameter ($\nu=0.1$), where the overall performance trends remain consistent. 
The peak throughput decreases from $148$ Mbps to $120$ Mbps, as expected, due to the reduced source entropy under a larger shaping parameter. 
The gains from the BGMD become more pronounced: With CB-SIC, BGMD provides a $2$--$5$\,dB SNR gain over UCD. 
This shows that exploiting PS priors in transceiver design yields more gains as the PS symbol distribution deviates from the uniform distribution.

\tcmajor{
To evaluate the transceiver under practical link adaptation, we introduce a PS-based MCS table (PS-MCS) as a counterpart to the 5G NR baseline (NR-MCS, Table 5.1.3.1-2 in \cite{3gpp38214}). 
Each PS-MCS index $m$ specifies a modulation order $Q_m$, an LDPC code rate $R_m$, and a shaping parameter $\nu_m$~\cite{hu2023supporting}. 
The shaping parameter $\nu_m$ is selected such that the resulting spectral efficiency $\mathrm{SE}_m^{PS}=2(H_{\nu}+1)-(1-R_m)Q_m$ matches that of the corresponding NR-MCS index  (i.e., $\mathrm{SE}_{m}^{NR}$), where $H_\nu=-\sum_{s\in\mathcal{A}}\mathbb{P}_{\nu}(s)\log_2{\mathbb{P}_{\nu}(s)}$ denotes the MB shaping entropy.  
For each target SE, the best combination $(Q_m, R_m, \nu_m)$ is obtained from extensive BLER simulations. 
% {Note that the best combination $(Q_m, R_m, \nu_m)$ of a given SE is obtained from extensive BLER simulations.}
MCS 0--3 are retained from NR-MCS with no PS ($\nu=0$), as PS offers negligible gains at low modulation orders and code rates. 
The modulation order is assigned as 16-QAM, 64-QAM, 256-QAM, and 1024-QAM for MCS 4--9, 10--16, 17--22, and 23--27, respectively.
The code rates are assigned as $R=2/3$, $3/4$, $4/5$, and $6/7$ for MCS 4--6, 7--12, 13--18, and 19--27, respectively.
The pair $(Q_m,R_m)$ is chosen to satisfy $R_mQ_m > \mathrm{SE}_{m}^{NR}$, ensuring a feasible $\nu_m>0$.
Link adaptation follows an outer-loop mechanism that selects the PS-MCS index to maintain a target BLER of $10\%$. 
Since each PS-MCS index specifies a shaping parameter, the transceiver can derive the BGMD precoder without any additional signaling overhead.
}

\begin{figure}[t]		\centering %\setcounter{figure}{0}
		\includegraphics[scale=0.5]{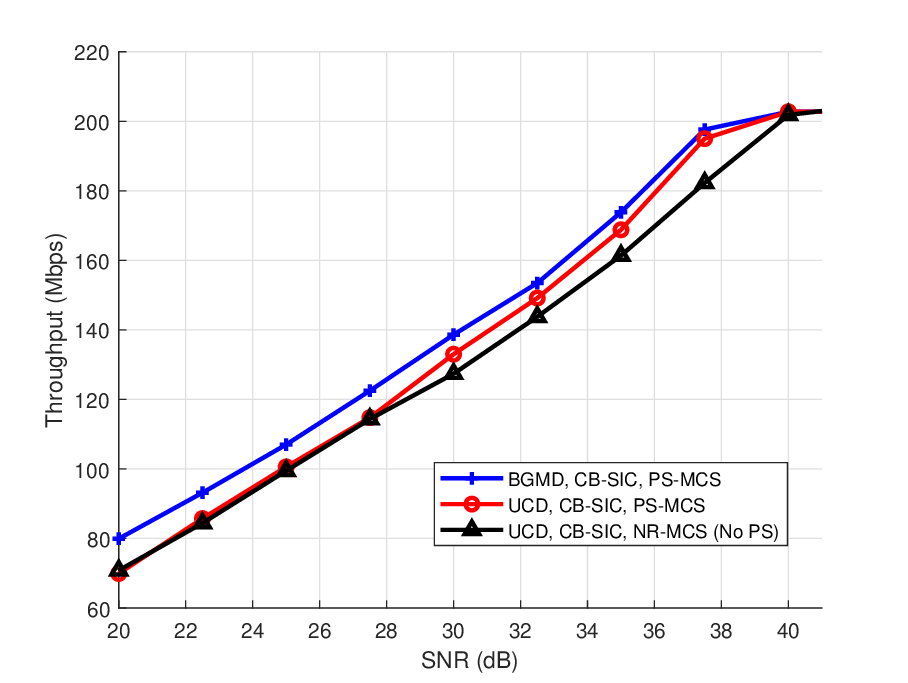}
		\caption{
            \tcmajor{Throughput versus SNR with link adaptation. }
			% \tcmajor{Throughput versus SNR, with Link Adaptation Results.}
		}
		\label{fig:LinkAdaptation_MCS_Comparison}
		% \vspace{-3mm}
\end{figure}

\tcmajor{Fig.~\ref{fig:LinkAdaptation_MCS_Comparison} evaluates the throughput under link adaptation with PS-MCS. Both BGMD and UCD employ MAP CB-SIC with PS-MCS, while the baseline UCD transceiver uses MMSE CB-SIC with NR-MCS. The PS-MCS schemes outperform the NR-MCS baseline, confirming the throughput benefits of PS. 
With PS-MCS, BGMD provides a 0.2--1.5\,dB SNR gain over UCD under link adaptation.
% Furthermore, BGMD achieves a throughput gain up to $1.5$ dB over UCD across SNR = 20dB - 40dB. 
% The gain becomes more pronounced at medium-to-high SNR, where the impact of the shaping-aware precoder is more significant. 
These results demonstrate that incorporating PS priors into the BGMD precoder design further improves throughput performance.}

\begin{figure}[t]		\centering %\setcounter{figure}{0}
		\includegraphics[scale=0.5]{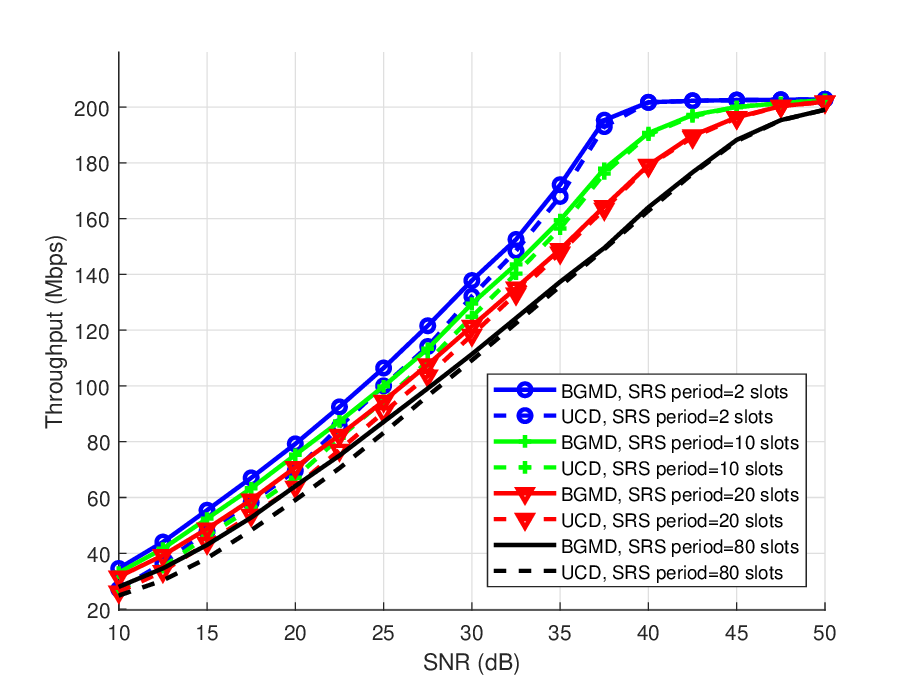}
        
		\caption{
            \tcmajor{Throughput versus SNR with channel aging for different SRS periods}.
			% \tcmajor{Channel aging impact on throught performance.}
		}
		\label{fig:ChannelAging_Comparison}
		% \vspace{-3mm}
\end{figure}

\tcmajor{Fig.~\ref{fig:ChannelAging_Comparison} evaluates the BGMD under channel aging for different sounding reference signal (SRS) periods, where the precoder is updated upon each SRS reception.
The link adaptation with PS-MCS is adopted.
% assuming that the estimated channel for precoder is updated when SRS is received. 
As the SRS period increases, the throughput of both BGMD and UCD degrades due to the mismatch between the channel used for precoder design and the true transmission channel. With frequent CSI updates  of 2-slot SRS period, BGMD provides up to 1.5~dB SNR gain over UCD. Although the gain gradually diminishes as channel aging becomes more severe, BGMD still outperforms UCD in the low-to-medium SNR regime even for an 80-slot SRS period. These results indicate that the BGMD remains beneficial under practical CSI imperfections.}

% \vspace{-2mm}
\section{Conclusion}
This work proposed a distribution-aware GMD transceiver design that incorporates PS priors to enhance throughput and decoding performance in MIMO transmission.
By leveraging the PS-induced symbol distribution, we derived an optimal MAP symbol detector and implemented it within a VBLAST detection framework.
% By leveraging the PS-induced symbol distribution, we derived an optimal MAP-VBLAST detection framework. 
We developed a BGMD precoder to exploit PS priors and equalize the per‑layer post-detection SNR, enabling the use of a single-MCS transmission with reduced complexity while preserving channel capacity.
% We developed a BGMD precoder exploitng the PS priors in the derivation and equalizes the per‑layer decoding SNR to facilitate the single modulation and coding scheme for low-complexity transmissions while preserving channel capacity.
In addition, we proposed a new CW-to-layer mapping scheme, LC-MIMO mapping, to enable CB-SIC and mitigate cross‑layer error propagation.
Numerical results showed that the BGMD transceiver with CB-SIC in LC-MIMO achieves significant gains in BLER and throughput compared with state-of-the-art transceivers when PS QAM symbols are employed.
% significantly improves decoding BLER and empirical throughput compared to state-of-the-art transceivers when PS QAM symbols are employed.

% QAM‑constrained achievable throughput and decoding performance compared to state‑of‑the‑art transceivers when PS‑QAM signaling is employed.

% by integrating PS priors to achieve higher data rates in MIMO transmission.
% By leveraging the PS symbol distribution, we derived an optimal MAP estimator, simplified with a VBLAST symbol detection framework.
% Then, we designed a BGMD precoder to equalize the per-layer decoding SNR while preserving the channel's maximum Gaussian mutual information.
% Furthermore, we proposed a new CB-to-layer mapping scheme, LC-MIMO mapping, to mitigate cross-layer error propagation in SIC.
% Numerical results showed that the BGMD transceiver with LC-MIMO improves QAM-constrained achievable throughput and decoding performance compared to the state-of-the-art transceivers when PS QAM symbols are employed.

% does achieve an enhanced decoding performance over the state-of-the-art transceiver schemes, when PS modulated symbols is enabled in MIMO transmissions.

% \vspace{-3mm}
\bibliographystyle{IEEEtran}
\bibliography{IEEEabrv,reference}

% Generated by IEEEtran.bst, version: 1.14 (2015/08/26)
\begin{thebibliography}{10}
\providecommand{\url}[1]{#1}
\csname url@samestyle\endcsname
\providecommand{\newblock}{\relax}
\providecommand{\bibinfo}[2]{#2}
\providecommand{\BIBentrySTDinterwordspacing}{\spaceskip=0pt\relax}
\providecommand{\BIBentryALTinterwordstretchfactor}{4}
\providecommand{\BIBentryALTinterwordspacing}{\spaceskip=\fontdimen2\font plus
\BIBentryALTinterwordstretchfactor\fontdimen3\font minus \fontdimen4\font\relax}
\providecommand{\BIBforeignlanguage}[2]{{%
\expandafter\ifx\csname l@#1\endcsname\relax
\typeout{** WARNING: IEEEtran.bst: No hyphenation pattern has been}%
\typeout{** loaded for the language `#1'. Using the pattern for}%
\typeout{** the default language instead.}%
\else
\language=\csname l@#1\endcsname
\fi
#2}}
\providecommand{\BIBdecl}{\relax}
\BIBdecl

\bibitem{brinton2025key}
C.~G. Brinton \emph{et~al.}, ``{Key focus areas and enabling technologies for 6G},'' \emph{IEEE Commun. Mag.}, vol.~63, no.~3, pp. 84--91, 2025.

\bibitem{hu2023supporting}
S.~Hu, H.~Wang, and S.~Semenov, ``{Supporting probabilistic constellation shaping in 5G-NR evolution},'' \emph{IEEE Trans. Wireless Commun.}, vol.~23, no.~4, pp. 3586--3599, 2023.

\bibitem{fang2025probabilistic}
J.~Fang, Q.~Li, C.~Chen, A.~Gurevitz, and Y.~Yoffe, ``{Probabilistic Shaping for Wi-Fi 8},'' \emph{IEEE J. Sel. Areas Commun.}, vol.~43, no.~11, pp. 3708--3721, 2025.

\bibitem{jiang2005joint}
Y.~Jiang, J.~Li, and W.~W. Hager, ``{Joint transceiver design for MIMO communications using geometric mean decomposition},'' \emph{IEEE Trans. Signal Process.}, vol.~53, no.~10, pp. 3791--3803, 2005.

\bibitem{jiang2005ucd}
------, ``{Uniform channel decomposition for MIMO communications},'' \emph{IEEE Trans. Signal Process.}, vol.~53, no.~11, pp. 4283--4294, 2005.

\bibitem{yang2021joint}
F.~Yang and Y.~Dong, ``{Joint probabilistic shaping and beamforming scheme for MISO VLC systems},'' \emph{IEEE Wireless Commun. Lett.}, vol.~11, no.~3, pp. 508--512, 2021.

\bibitem{kang2022probabilistic}
W.~Kang, ``{A Probabilistic Shaping Scheme for MIMO Systems with Signal Space Diversity},'' in \emph{2022 IEEE Wireless Communications and Networking Conference (WCNC)}.\hskip 1em plus 0.5em minus 0.4em\relax IEEE, 2022, pp. 251--255.

\bibitem{ivanov2025probabilistic}
K.~Ivanov, W.~Yang, and J.~Jiang, ``{Probabilistic Shaping in MIMO: Going Beyond 1.53dB AWGN Gain With the Non-Linear Demapper},'' in \emph{Proc. 13th Int. Symp. Topics in Coding (ISTC)}, 2025, pp. 1--5.

\bibitem{studer2008soft}
C.~Studer, A.~Burg, and H.~Bolcskei, ``{Soft-output sphere decoding: Algorithms and VLSI implementation},'' \emph{IEEE J. Sel. Areas Commun.}, vol.~26, no.~2, pp. 290--300, 2008.

\bibitem{sadek2007leakage}
M.~Sadek, A.~Tarighat, and A.~H. Sayed, ``{A leakage-based precoding scheme for downlink multi-user MIMO channels},'' \emph{IEEE Trans. Wireless Commun.}, vol.~6, no.~5, pp. 1711--1721, 2007.

\bibitem{cheng2010new}
P.~Cheng, M.~Tao, and W.~Zhang, ``{A new SLNR-based linear precoding for downlink multi-user multi-stream MIMO systems},'' \emph{IEEE Commun. Lett.}, vol.~14, no.~11, pp. 1008--1010, 2010.

\bibitem{bocherer2015bandwidth}
G.~B{\"o}cherer, F.~Steiner, and P.~Schulte, ``{Bandwidth efficient and rate-matched low-density parity-check coded modulation},'' \emph{IEEE Trans. Commun.}, vol.~63, no.~12, pp. 4651--4665, 2015.

\bibitem{schulte2015constant}
P.~Schulte and G.~B{\"o}cherer, ``{Constant composition distribution matching},'' \emph{IEEE Trans. Inf. Theory}, vol.~62, no.~1, pp. 430--434, 2015.

\bibitem{gultekin2019enumerative}
Y.~C. G{\"u}ltekin, W.~J. van Houtum, A.~G. Koppelaar, and F.~M. Willems, ``Enumerative sphere shaping for wireless communications with short packets,'' \emph{IEEE Trans. Wireless Commun.}, vol.~19, no.~2, pp. 1098--1112, 2019.

\bibitem{kay1993fundamentals}
S.~M. Kay, \emph{Fundamentals of Statistical Signal Processing: Estimation Theory}.\hskip 1em plus 0.5em minus 0.4em\relax USA: Prentice-Hall, Inc., 1993.

\bibitem{hassibi2000efficient}
B.~Hassibi, ``{An efficient square-root algorithm for BLAST},'' in \emph{Proc. Int. Conf. ASSP}, vol.~2.\hskip 1em plus 0.5em minus 0.4em\relax IEEE, 2000, pp. 737--740.

\bibitem{3gpp38211}
3GPP, ``{NR; Physical channels and modulation},'' Technical Specification (TS) 38.211, 02 2026, version 19.2.0.

\bibitem{3gpp38901}
------, ``Study on channel model for frequencies from 0.5 to 100 {GHz},'' Technical Report (TR) 38.901, 01 2024, version 17.1.0.

\bibitem{3gpp38212}
------, ``{NR; Multiplexing and channel coding},'' Technical Specification (TS) 38.212, 02 2026, version 19.2.0.

\bibitem{3gpp38214}
------, ``{NR; Physical layer procedures for data},'' Technical Specification (TS) 38.214, 06 2026, version 19.4.0.

\end{thebibliography}

\end{document}